\documentclass[amsmath,reprint,showkeys,nofootinbib]{revtex4-1}
\usepackage[latin1]{inputenc}
\usepackage{amsfonts,amssymb}
\usepackage{graphicx} 
\usepackage{bbm}
\usepackage{bm}
\renewcommand{\P}{\mathcal{P}}

\def\id{\ensuremath{\mathbbm{1}}}

\def\Tr{\ensuremath{\mathrm{Tr}}}
\usepackage[hypertexnames=false]{hyperref}
\begin{document}
\title{Postselection induced entanglement swapping from a vacuum--excitation entangled state 
to separate quantum systems}
\author{Antonio Di Lorenzo}
\affiliation{Instituto de Física, Universidade Federal de Uberlândia, Av. João Naves de Ávila 2121, 
Uberlândia, Minas Gerais, 38400-902,  Brazil}%
\affiliation{CNR-IMM-UOS Catania (Università), Consiglio Nazionale delle Ricerche,
Via Santa Sofia 64, 95123 Catania, Italy}
\begin{abstract}
We show that a single particle in a superposition of different paths can entangle two objects located on each path. 
The entanglement has its maximum visibility for intermediate coupling strengths. 
In particular, when the two quantum systems with which the particle interacts 
are detectors that measure its presence and its polarization, the so-called quantum Cheshire cat is realized. 
\end{abstract}
\keywords{Entanglement, Quantum inseparability, Quantum paradoxes}
\pacs{03.65.Ta,03.65.Ud,03.67.Bg,03.67.Mn}
\maketitle
The two most perplexing features of quantum mechanics are the interference in a double slit experiment, and the entanglement of spatially separated systems. 
In the paradigmatic double slit experiment, a particle, in some sense, follows two paths at the same time, as 
shown by the appearance of interference fringes after accumulating many measurements. 
This feature is very elusive, as trying to measure the presence of the particle on either path 
destroys the interference. 
A strong evidence in favor of this ubiquity is that a single particle can induce entanglement in two separated quantum systems that have never mutually interacted and that are placed each on a different path, as if it interacted with both systems simultaneously. 
The preceding literature on this topic considered only the case of a strong interaction. 
A recent related proposal of Aharonov \emph{et al.} \cite{Aharonov2013}, where the effect is dubbed a quantum Cheshire cat, on the other hand, 
considers the weak coupling limit. 
Here, we tackle the problem for an arbitrary coupling. 
In the following we shall demonstrate that  a particle in a coherent 
superposition of spatially separated paths can induce entanglement between two distant meters located one on each path, 
and we propose an entanglement indicator to quantify it. 
We also demonstrate that the optimal couplings are not weak, but either strong or intermediate, depending 
on the dimensions of the Hilbert spaces of the meters.

Measurements can be divided approximately into three categories: strong, intermediate, and weak. 
Strong measurements are the textbook measurements that are described already in von Neumann book \cite{vonNeumann1932}. 
If a meter in a state $|A_0\rangle$ interacts with a quantum system prepared in an eigenstate $|O\rangle$ of the 
observable $\Hat{O}$ to be measured, the joint state of the two after the interaction is $|A_O,O\rangle$, where 
$|A_O\rangle$ are mutually orthogonal states of the meter. 
The evolution for an arbitrary initial state $|\psi\rangle$ of the system follows from the linearity of quantum mechanics.  
In weak measurements \cite{Aharonov1988}, the final states of the meter, $|A_O\rangle$, instead, are almost indistinguishable, 
$|A_O\rangle\simeq N_O ( |\bar{A}\rangle + g f(O) |\delta_O\rangle)$, with $g$ an effective coupling constant, 
$f$ a function of $O$, $|\delta_O\rangle$ a set of states not necessarily distinct, and $N_O$ a normalization.  
Perhaps ``measurement'' is a misleading term, since due to the weak interaction between the system and the meter 
one cannot infer substantial information about the former from a single trial. However, the statistical analysis of the postselected data --- 
which generally is limited to the average readout but could be extended to the full statistics 
\cite{DiLorenzo2012a,DiLorenzo2012e,Struebi2013,Ferrie2014} ---
allows to extract information about the system that is not trivially recovered from standard strong projective measurements. 
For instance, weak measurements followed by postselection  provide a powerful inference technique, allowing e.g. to reconstruct the unknown wavefunction of a system \cite{Lundeen2011}, or the density matrix \cite{Hofmann2010,Lundeen2012,Wu2013}. 
The coherent quantum nature of the meter was shown to be of the essence for the peculiar 
amplification of the weak measurement \cite{Duck1989,DiLorenzo2008}. Several experimental works 
have focused on signal amplification \cite{Ritchie1991,Hosten2008,Dixon2009,Gorodetski2012}, but the efficiency of the amplification has been questioned 
due to the corresponding decrease in the probability of postselection \cite{Tanaka2013,Knee2013a,Knee2014,Ferrie2014}. 
In intermediate measurements, on the other hand, no special form for the output states $|A_O\rangle$ is postulated, 
but the evolution $U_g |A_0,O\rangle$ is calculated by assuming a sensible form for the interaction, depending on a parameter $g$. 
For $g\to\infty$ and for $g\to0$ the strong and the weak measurement limit are recovered.  
Intermediate couplings can 
perform at least as well as weak couplings for determining the unknown state of a system by 
sequential measurements \cite{DiLorenzo2011a,DiLorenzo2013a,DiLorenzo2013f}, and they can also be used 
for obtaining a violation of the original Heisenberg inequality 
for momentum and position in the noise-disturbance formulation \cite{DiLorenzo2013c}. 
While both the weak and the strong measurements can be treated in a mathematically simple way, as they are ideal limiting cases, 
the intermediate regime often requires a numerical approach, but analytical results can be obtained by making the simplifying 
assumptions of a nondemolition interaction and of an initial Gaussian state for the meter.   
In the case studied in the rest of this Letter, the intermediate measurement will prove to be superior both to the strong and 
to the weak measurement. 
%

\begin{figure}
\centering
\includegraphics[width=3in]{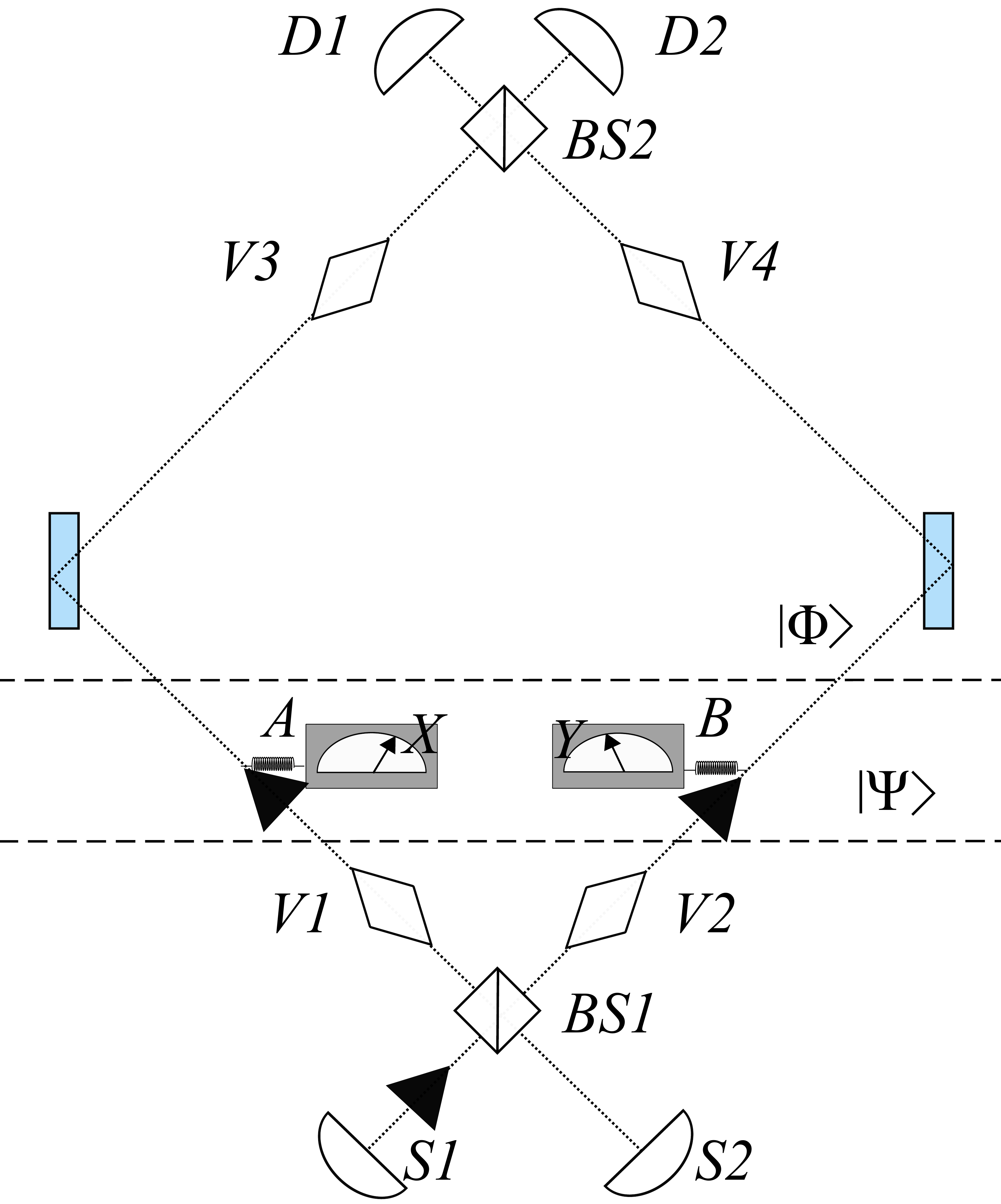}
\caption{\label{fig:setup} The setup is a variant of the Mach-Zehnder interferometer. The source $S1$ emits a polarized particle which enters a beam-splitter $BS1$, 
and exits in a coherent superposition of spatially separated states. 
In a sense, the particle is simultaneously in the left and in the right arm, as it can induce entanglement between 
two quantum systems $A$ and $B$, as can be evinced by measuring the cross-moment between two observables $X$ and $Y$. The entanglement can be observed conditionally on the postselection, made by a judicious combination of a second beam-splitter and of polarization-sensitive detectors $D_j$. The local unitary operators $V_j$ allow to arbitrarily set the preparation $\Psi$ and the post-selection $\Phi$,}
\end{figure}
%
Let us discuss the simple case of preparation $|\Psi\rangle$ and postselection $|\Phi\rangle$ in a pure state 
for the system, and of an initial pure (and uncorrelated) state for the meters $|A_0,B_0\rangle$. 
In a nondemolition measurement, 
if the photon is in the left arm, the total state evolves to 
$|L,\sigma,A_1,B_0\rangle$; if instead the photon is in the right arm and it has positive (negative) polarization, $|\Psi\rangle=|R,\pm\rangle$, 
the final state is $|R,\pm,A_0,B_\pm\rangle$. 
Because of the superposition principle, if the system is in $|\Psi\rangle=a|L,\sigma\rangle+b|R,+\rangle+c|R,-\rangle$, 
the final state is $a|L,\sigma,A_1,B_0\rangle+b|R,+,A_0,B_+\rangle+c|R,-,A_0,B_-\rangle$. 
Now, if the photon is traced out, the final state of the meters is a mixture, 
$\rho_{cl.corr.}=|a|^2 |A_1,B_0\rangle\langle A_1,B_0|+|b|^2 |A_0,B_+\rangle\langle A_0,B_+|+|c|^2 |A_0,B_-\rangle\langle A_0,B_-|$, that shows only classical correlations. 
If instead the photon is successfully postselected in a state $|\Phi\rangle$, the final state of the meters is entangled, as 
\begin{equation}
|F\rangle = l|A_1,B_0\rangle
+r^+|A_0,B_+\rangle+r^-|A_0,B_-\rangle,
\label{eq:pure}
\end{equation} 
where we defined the complex transition amplitudes 
\begin{equation}
 l = \langle \Phi|\Pi_L|\Psi\rangle , \qquad
r^\pm= \langle \Phi|\Pi_{R,\pm}|\Psi\rangle,
\end{equation}
with $\Pi_L=\sum_\pm|L,\pm\rangle\langle L,\pm|$ the rank-2 projector in the left arm, and 
$\Pi_{R,\pm}=|R,\pm\rangle\langle R,\pm|$ the rank-1 projectors on the right arm with polarization $\pm$ 
\footnote{We note that, because of the completeness relation \unexpanded{$\Pi_L+\Pi_{R+}+\Pi_{R-}=\id$}, 
\unexpanded{$l+r^++r^-=\langle\Phi|\Psi\rangle$}. It is customary to define 
the weak values \unexpanded{$L_{w}=l/(l+r^++r^-)$} and 
\unexpanded{$\Sigma_{w}=(r^+-r^-)/(l+r^++r^-)$}, 
associated, respectively, to the operators \unexpanded{$\Pi_{L}$} and \unexpanded{$\sigma_{R}=\Pi_{R,+}-\Pi_{R,-}$}. 
However, we prefer to use transition amplitudes, which are always well behaved.}.  
The state $|F\rangle$ in \eqref{eq:pure} is not normalized to one, instead $\langle F|F\rangle=\P$, 
the probability of a successful postselection. 
If the postselection fails, the unnormalized final state of the meters is mixed, $\check{\rho}_f=\rho_{cl.corr.}-|F\rangle\langle F|$. 
Notice how the trace $\Tr(\check{\rho}_f)=\P'=1-\P$ is the probability for the postselection to fail. 

The entanglement  is due to the photon being in a coherent superposition of states localized in the left and in the right arm, 
so that, in some sense, it interacts with both meters at the same time. 
If no postselection would occur, or, more generally, if the preparation or the postselected state would not be 
a coherent superposition of states localized in the left and in the right arm, the entanglement would not manifest. 
The situation is analogous to delayed--choice entanglement swapping \cite{Peres2000}, 
but here there are only three quantum systems (the particle and the two pointers), and no 
preexisting entanglement among them seems to be present. 
As a matter of fact, however, we are in presence of vacuum--excitation entanglement \cite{Tan1991,Hardy1994,Lombardi2002,Sciarrino2002}, which is swapped to the meters. 
Indeed, a superposition of a photon in the left and the right arm can be written as 
$a|1_{L,\sigma},0_{L,-\sigma},0_{R,+},0_{R,-}\rangle+b|0_{L,\sigma},0_{L,-\sigma},1_{R,+},0_{R,-}\rangle+c|
0_{L,\sigma},0_{L,-\sigma},0_{R,+},1_{R,-}\rangle$, having the vacuum state of the electromagnetic field present in the 
left or right propagating channel, and its excitation, the photon, present in the right or left propagating channel.  The fact that the particle is a photon is irrelevant (we are calling it a photon just to fix the ideas, indeed), the same rationale applies to any other particle, which can be considered an excitation of a quantum field. 
The issue of whether a single particle is actually entangled is quite debated \cite{Pawlowski2006,Drezet2006,Enk2006}, 
and we shall not address it here. We are content with the uncontroversial fact that the two distant meters get entangled without 
having interacted, and we shall not debate whether this entanglement was swapped from the single-particle entanglement or 
whether it was created by a nonlocal interaction. 

How to detect the entanglement between the meters? 
Let us consider an unnormalized average of the form $m=\langle F|\Hat{X}_A\Hat{X}_B|F\rangle$, 
with $\Hat{X}_A$ an observable of the meter $A$ and $\Hat{X}_B$ an observable of the meter $B$. 
We have $m=m_{cl}+m_{ent}+m_{l.i.}$, where 
\begin{align}
m_{cl}&= |l|^2  \langle A_1 |\Hat{X}_A|A_1\rangle \langle B_0 |\Hat{X}_B|B_0\rangle 
\nonumber\\
&\qquad+\sum_\pm|r^\pm|^2 \langle A_0|\Hat{X}_A|A_0\rangle \langle B_\pm|\Hat{X}_B|B_\pm\rangle
\end{align}
is the classical part, 
\begin{equation}
m_{ent}= \sum_\pm 
2\Re\left(l^* r^\pm \langle A_1 |\Hat{X}_A|A_0\rangle \langle B_0 |\Hat{X}_B|B_\pm\rangle\right) 
\end{equation}
is the contribution from the interference between the two meters, i.e. from their entanglement, and 
\begin{equation}
m_{l.i.}=  
2\Re\left(r^{+*} r^- \langle A_0 |\Hat{X}_A|A_0\rangle \langle B_+|\Hat{X}_B|B_-\rangle\right)
\end{equation}
is the contribution from the local interference in the meter $B$. If either $\Hat{X}_A$ or $\Hat{X}_B$ is the identity, 
in the strong coupling limit, the contribution from the entanglement vanishes. Therefore, one needs to consider 
two nontrivial operators, as in the case of Bell inequalities. In the weak coupling limit, however, entanglement contributes 
to the average $m$ even if it is a local average, because $\langle A_1|A_0\rangle\simeq \langle B_\pm|B_0\rangle\simeq 1$.  

The strong interaction limit was considered in the former literature \cite{Gerry1996,Enk2005,Ashhab2007}. 
The recent proposal of Aharonov \emph{et al.} \cite{Aharonov2013}, instead, adopts the same scheme 
(the authors are apparently unaware of this), but with a weak coupling. 

If it is possible to make a measurement on the meters that  
 projects their states into arbitrary combinations $\alpha|A_0\rangle+\beta|A_1\rangle$ and 
$\alpha|B_0\rangle+\beta|B_+\rangle+\gamma|B_-\rangle$, then one could check the violation of a Bell-like inequality \cite{Bell1964,Clauser1969,Clauser1974}, or, better, one could use the criteria discussed by Peres \cite{Peres1996} and Horodecki \emph{et al.} \cite{Horodecki1996}, as the entanglement is between a two-level system and a three-level system. 
In this case, the maximum entanglement is achieved for a strong interaction, so that $\{|A_0\rangle,|A_1\rangle\}$ 
and $\{|B_0\rangle,|B_\pm\rangle\}$ form orthogonal bases. 
Thus, we have reached a first partial conclusion: if the meters have a finite-dimensional Hilbert space, whose relevant two-- and three--dimensional subspaces can be probed projectively along any basis, then it is possible to observe the entanglement 
induced by the postselection already in the strong coupling regime. 

However, if the meters have an infinite dimensional Hilbert space, 
the task of making projective measurements on $\alpha|A_0\rangle+\beta|A_1\rangle$ and 
$\alpha|B_0\rangle+\beta|B_+\rangle+\gamma|B_-\rangle$ may be a practical impossibility. 
Furthermore, unwanted external influences can drive the states of the meters away from the simple two-- and three--dimensional 
subspaces spanned by these bases. 
Therefore another criterion for entanglement should be used. 
Our goal is to find observables $\hat{o}_A$ and $\hat{o}_B$ such that $m_{cl}=m_{l.i}=0$ and $m_{ent}\neq 0$, 
so that $m$ works as an unambiguous indicator for entanglement. A sufficient condition is that 
$\langle A_0 |\Hat{X}_A|A_0\rangle= \langle B_0 |\Hat{X}_B|B_0\rangle=0$, i.e. if the particle is not in the left (respectively, right) 
arm, the expectation value of $\Hat{X}_A$ (resp., $\Hat{X}_B$) is zero. We note, however, that 
the observed value of $X_A$ may not be zero, as we are not requiring a strong measurement  ---which implies 
that the state $|A_0\rangle$ is an eigenstate of $\Hat{X}_A$ with null eigenvalue--- thus quantum statistical fluctuations and environmental noise can yield a nonzero result in an individual trial. 

We indicate with $x$ and $y$ the pointers of the meters, whose initial states have the representation $\langle x|A_0\rangle=\phi_0(x)$, $\langle y|B_0\rangle=\phi_0(y)$.  
By pointers, we mean that, in the strong coupling regime, observing $x$ and $y$ gives unambiguous information 
about the presence of the particle in the left arm and the value of its polarization in the right arm. 
Think of the Stern-Gerlach apparatus, where the position of a spot on the screen is the pointer revealing the value of the 
spin of the atom. 
The meters are assumed to be unbiased, so that the initial averages of the pointers $x$ and $y$ with $|\phi_0(x)\phi_0(y)|^2$ 
are null. We shall consider the pointers in units of the initial uncertainties $\Delta_{A}$ and $\Delta_{B}$, i.e. 
$\int\!dx\, x^2|\phi_0(x)|^2=\int\!dy\, y^2|\phi_0(y)|^2=1$. 
For simplicity, we assume the von Neumann model of measurement. 
In this model, after the interaction with a particle, the wave functions of the meters become 
$\langle x|A_1\rangle=\phi_0(x-g_{A})$,  $\langle y|B_\pm\rangle=\phi_0(y\mp g_{B})$, 
with $g_{A},g_{B}$ the dimensionless coupling constants. 
The final state of the meters, in the pointers representation, is 
$\langle x,y|F\rangle = \phi(x,y)$. 

In optics, it is possible to realize arbitrary couplings $g_A, g_B$, respectively, by using a refractive crystal 
that dislocates the beam along the $x$ axis by an amount $\delta x = g_A \Delta_x$, 
and a birefringent crystal of appropriate length, so that the beams exiting the latter have a separation $\delta y = g_B \Delta_y$, 
with $\Delta_x, \Delta_y$ the variance of the input beam. 

As the entanglement indicator, we shall consider the cross-moment 
$\langle xy\rangle \P= \langle F|\Hat{x}\Hat{y}|F\rangle$, 
to which only the entanglement terms give a nonzero contribution, 
\begin{align}
\langle x y\rangle \P&=
2\sum_\pm\Re\biggl[
l^* r^\pm
\int\!dx\, x \phi_0(x)\phi^*_0(x-g_{A})
\nonumber
\\
& \qquad\times
\int\!dy\, y\phi_0(y\mp g_{B})\phi^*_0(y)\biggr].
\label{eq:avcross}
\end{align}
It may happen that the two contributions from the $\phi_0(y+g_B)$ and the $\phi_0(y-g_B)$ wave function 
are present but they cancel out. 
In this case, entanglement may be detected by using another cross-moment,  as we shall discuss elsewhere \cite{DiLorenzo2014}. 

When the postselection fails, which happens with probability $\P'=1-\P$, 
the entanglement indicator is $\langle x y\rangle_{f} \P'=\Tr(\Hat{x}\hat{y}\check{\rho}_{f})=
\Tr(\Hat{x}\hat{y}\rho_{cl.corr})- \langle F|\Hat{x}\Hat{y}|F\rangle=-\langle x y\rangle \P$, 
where we used the fact that $\Tr(\Hat{x}\hat{y}\rho_{cl.corr})=0$, i.e., there is no entanglement contribution to classical 
correlations. 
Therefore, we can use all the experimental data by defining the entanglement indicator as follows: 
In the $j$-th trial, if the postselection is successful, consider the product $c_j=x_j y_j$, 
otherwise, consider $c_j=-x_j y_j$; sum the $c_j$ and divide by the number of trials; 
the value $\mathcal{C}=2\langle x y\rangle \P$ is thus obtained, 
allowing to establish whether a Cheshire cat is observed ($\mathcal{C}\neq 0$) or not 
($\mathcal{C}= 0$). 
Formally, if we assign a binary variable $\tau=\pm 1$ to the postselection of the photon, with $\tau=+1$ representing 
a successful postselection in $E$, and with $\tau=-1$ representing a failed postselection, 
the Cheshire cat parameter is given by the signed cross-moment 
 \begin{equation}
\mathcal{C}=\langle \tau xy\rangle_a ,
\label{eq:expc}
\end{equation}
which provides the signature 
for the entanglement between the meter measuring the presence of the photon 
and the meter measuring its polarization in the two arms of the interferometer. 
The index $a$ is a reminder that the average in \eqref{eq:expc} is made over all the experimental data, not only on 
the ones obtained for a successful postselection. 
 
So far, we have provided exact results. 
In the weak coupling limit, a shift of $g_{A}$ is small compared to the range over which $\phi_0$ varies appreciably, 
so that one can approximate $\phi_0(x-g_{A})\simeq \phi_0(x)-g_{A}d\phi_0(x)/dx$, etc. As $id/dx$ represents the momentum operator, it is possible to approximate the overlap integrals with appropriate combinations of the initial averages of $\hat{x}\hat{p}_x$, etc. \cite{Jozsa2007} .  
In the strong coupling limit, instead, the wave functions $\phi_0(x)$ and $\phi_0(x-g_{A})$ have a negligible overlap, 
$\phi_0^*(x-g_{A})\phi_0(x)\simeq 0$, etc., so that the interference terms disappear. 
Precisely, the overlap terms of interest in \eqref{eq:avcross} behave asymptotically as 
\begin{subequations}
\begin{align}
&\left|\int\!dx\, x \phi_0^*(x)\phi_0(x-g_{A})\right|\approx
\begin{cases}
g_{A}&\text{for } g_{A}\ll1\\
g_{A}|\phi_0(g_{A})|&\text{for } g_{A}\gg1
\end{cases}
\\
&\left|\int\!dy\, y \phi_0^*(y)\phi_0(y\mp g_{B})\right|\approx
\begin{cases}
g_{B}&\text{for } g_{B}\ll1\\
g_{B}|\phi_0(\pm g_{B})|&\text{for } g_{B}\gg1
\end{cases}
\end{align}
\end{subequations}
We make the following, fundamental consideration: 
\emph{The entanglement indicator $\mathcal{C}$ vanishes bi-linearly in the couplings 
$g_{A} g_{B}$ for a weak measurement, 
and it vanishes as $g_{A}g_{B}|\phi_0(g_{A})||\phi_0(g_{B})|$ for a strong measurement. Therefore, there must be an optimal intermediate coupling strength 
for which the entanglement is not only present, but it gives a maximum contribution to $\mathcal{C}$.}

Thus, we need an expression working for any coupling strength, in order to determine the optimal one. 
We shall  consider the initial state of the meters to be Gaussian, $\phi_0(x)\propto \exp(-x^2/4)$, 
so that the overlap integrals can be calculated analytically. 
While \eqref{eq:expc} is the operational definition of the Cheshire cat parameter, as it can be obtained 
directly from experimental data, for Gaussian meters the exact theoretical value is 
\begin{equation}
\mathcal{C}=g_{A}g_{B} w_{A}w_{B}
\Re\left[\Tr{(E\sigma_{R}\rho\Pi_{L})}\right],
\label{eq:thc}
\end{equation}
with $E$ a mixed postselection state, $\rho$ a mixed preparation state, 
$\sigma_{R}=\sum_\pm \pm|R,\pm\rangle\langle R,\pm|$ the local spin operator in the right path, and  
$w_{A} = \exp{\left(-g_{A}^2/8\right)}$, $w_{B} = \exp{\left(-g_{B}^2/8\right)}$.
As a function of the preparation and the postselection, the extremal values of the Cheshire cat parameter is $\mathcal{C}_{max}=
g_{A}g_{B}w_{A}w_{B}/4$. 

More importantly, $|\mathcal{C}|$ is a non-monotonous function of the coupling constants. As we noted earlier, it goes to 
zero both in the weak coupling limit $g_{A,B}\to 0$, and in the strong 
coupling limit $g_{A,B}\to \infty$. Its extremal value, as a function of the couplings, 
is reached for 
$g_{A}= g_{B}=2$,
yielding 
$\mathcal{C}_{extr}=4 e^{-1}
\Re\left[\Tr{(E\sigma_{R}\rho\Pi_{L})}\right]$,
which, as a function of the preparation and postselection, has an absolute maximum 
$\mathcal{C}_{extr}=e^{-1}$. 
Therefore, the criterion $\mathcal{C}\neq 0$ does not require a very weak coupling, 
but it reaches its optimum when the coupling strengths are twice the initial uncertainty of the meters. 
Hence the optimal measurement is neither strong nor weak, but intermediate. 

We stress that we have so far assumed that 
the readout of the meters is projective and errorless, the only uncertainty $\Delta_{A},\Delta_{B}$ coming 
from the initial preparation of the meters. 
When external noise is accounted for, let 
us call its square variance $\nu_{A},\nu_{B}$, the criterion to observe unambiguously a Cheshire cat is that 
$\nu_{A}\nu_{B}\ll \mathcal{C}$. A necessary condition is that $\nu_{A}\ll \Delta_{A}$ and $\nu_{B}\ll \Delta_{B}$, i.e. 
the resolution of the readout must be much smaller than the initial uncertainty. 
By using once more the analogy with the Stern-Gerlach apparatus, this means that the input beam 
of silver atoms may have a waist $\Delta\gg g$, where $g$ is the deflection due to the magnetic field gradient, 
but the size of the spot on the screen created by each individual atom should be $\nu\ll\Delta$.

We conclude by comparing our criterion to the one used in Ref.~\cite{Aharonov2013}. 
The preparation and postselection were chosen by the authors of 
 Ref.~\cite{Aharonov2013} in such a way that, in the weak coupling limit, 
the average outputs take the special values 
$\lim_{g_{A}\to 0}\langle x/g_{A}\rangle =1$ and $\lim_{g_{B}\to 0}\langle y/g_{B}\rangle =1$. 
From this it was inferred that the photon is in the left arm, while its polarization is in the right arm. This phenomenon is called a quantum Cheshire cat, in the sense that a physical property can be separated from its carrier.  
The interpretation attributing to the averages $\lim_{g_{A}\to 0}\langle x/g_{A}\rangle =1$ and $\lim_{g_{B}\to 0}\langle y/g_{B}\rangle =1$ the meaning of having one particle on one path and its polarization on the other path is problematic. 
Indeed, it has been established since a long time\cite{Duck1989} that the averages $\langle x\rangle$ and $\langle y\rangle$ 
should not be interpreted literally as representing a value of the measured observable of the system.  
One should give these averages no more meaning than they have: they represent the average positions of pointers that have interacted with a quantum system. Their statistics differs from the classical statistics because of their own quantum nature, 
which leads to interference. In the present case, the interference is between two spatially separated meters, i.e. it manifests as entanglement. 
While in the weak coupling limit, as discussed above, entanglement does contribute to the local averages
$\langle x\rangle$ and $\langle y\rangle$, it is very difficult to unscramble the entanglement contribution in the latter two quantities. 
By contrast, the quantity $\mathcal{C}$ proposed here comes exclusively from the entanglement, and it is well defined 
for any coupling strength.  As such, it is better suited to characterize the presence of quantum correlations between the meters. 

In conclusion, we have demonstrated that the two fundamental aspects 
of quantum mechanics, coherence and entanglement, concur in the variant of the Mach--Zehnder interferometer proposed by Ref.~\cite{Aharonov2013}. 
The phenomenon seems to confirm the point of view that a 
single particle can be entangled with the vacuum, as separate quantum systems get entangled by interacting simultaneously with the  single particle. 

\begin{acknowledgments}
I am indebted to Alessandro Romito, Giuseppe Falci, Kavan Modi, Lugi Amico, and Philip Walther for discussions.
This work was performed as part of the Brazilian Instituto Nacional de Ci\^{e}ncia e
Tecnologia para a Informa\c{c}\~{a}o Qu\^{a}ntica (INCT--IQ), it
was supported by Funda\c{c}\~{a}o de Amparo \`{a} Pesquisa do 
Estado de Minas Gerais through Process No. APQ-02804-10 and 
by the Conselho Nacional de Desenvolvimento Cient\'{\i}fico e Tecnol\'{o}gico (CNPq) 
through Process no. 245952/2012-8. 
\end{acknowledgments}
\end{document}